\documentclass[sigconf]{acmart}

\usepackage{stfloats}

\copyrightyear{2026}
\acmYear{2026}
\setcopyright{cc}
\setcctype{by}
\acmConference[ICMI '26]{INTERNATIONAL CONFERENCE ON MULTIMODAL INTERACTION}{October 05--09, 2026}{Napoli, Italy}
\acmBooktitle{INTERNATIONAL CONFERENCE ON MULTIMODAL INTERACTION (ICMI '26), October 05--09, 2026, Napoli, Italy}
\acmDOI{10.1145/3776574.3831184}
\acmISBN{979-8-4007-2318-6/26/10}

\begin{document}

\title{Multimodal Rapport Estimation in Real-World HRI}

\author{Akihiro Sakuramoto}
\affiliation{%
  \institution{Japan Advanced Institute of Science and Technology}
  \city{Nomi}
  \state{Ishikawa}
  \country{Japan}
}
\email{s2510069@jaist.ac.jp}

\author{Takato Hayashi}
\affiliation{%
  \institution{Japan Advanced Institute of Science and Technology}
  \city{Nomi}
  \state{Ishikawa}
  \country{Japan}
}
\email{hayashi0884@jaist.ac.jp}

\author{Ryo Miyoshi}
\affiliation{%
  \institution{CyberAgent}
  \city{Tokyo}
  \country{Japan}
}
\affiliation{%
  \institution{The University of Osaka}
  \city{Osaka}
  \country{Japan}
}
\email{miyoshi\_ryo@cyberagent.co.jp}

\author{Yuki Okafuji}
\affiliation{%
  \institution{CyberAgent}
  \city{Tokyo}
  \country{Japan}
}
\affiliation{%
  \institution{The University of Osaka}
  \city{Osaka}
  \country{Japan}
}
\email{okafuji\_yuki\_xd@cyberagent.co.jp}

\author{Shogo Okada}
\affiliation{%
  \institution{Japan Advanced Institute of Science and Technology}
  \city{Nomi}
  \state{Ishikawa}
  \country{Japan}
}
\email{okada-s@jaist.ac.jp}

\renewcommand{\shortauthors}{Sakuramoto et al.}

\begin{abstract}
Evaluating interaction quality in real-world HRI is an important challenge. If interaction quality can be estimated reliably, the results can be used to improve dialogue strategies and ultimately enable robots to adapt their behavior autonomously. However, existing automatic evaluation methods have been developed primarily in controlled laboratory settings, and it remains unclear whether they can be directly applied to real-world environments, where users are free to disengage and multi-party participation may arise naturally. In this study, we investigate the automatic estimation of third-party-rated rapport scores using 62 sessions of multimodal recordings collected in a Japanese drugstore. We compare zero-shot LLMs, pretrained text, audio, and visual models, and their prediction-level fusion. The results show that, in real-world HRI, zero-shot LLMs achieve strong performance, while audio and visual models tend to provide complementary information. In particular, Gemini 2.5 Flash performs strongly as a single model, and a fusion model combining Gemini (text) with HuBERT and V-JEPA performs best overall. Further analyses showed that estimation performance varied across interaction-duration and group-size conditions. These findings suggest that rapport estimation in real-world HRI requires evaluation and model design that account for contextual variability beyond that assumed in laboratory settings.
\end{abstract}



\begin{teaserfigure}
  \includegraphics[width=\textwidth]{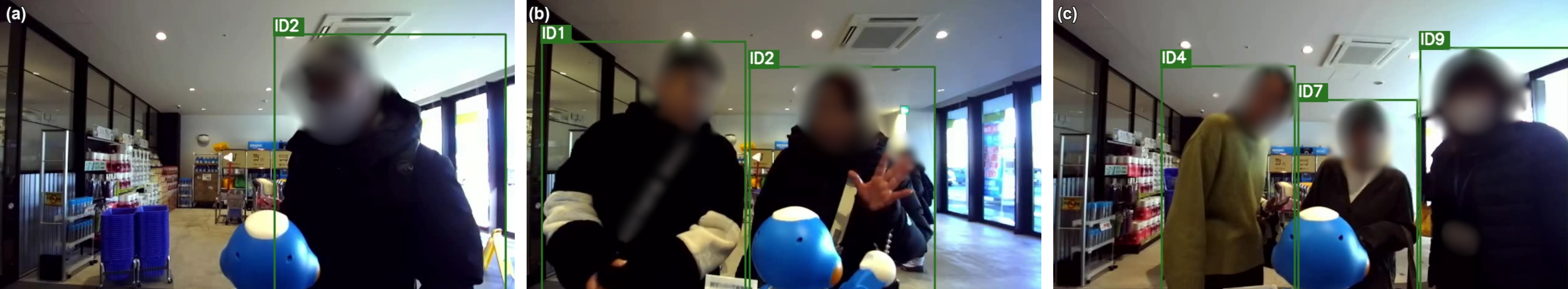}
  \caption{Robot-view examples from the real-world HRI corpus: (a) a single-participant interaction, (b) a two-participant interaction, and (c) a three-participant interaction.}
  \Description{Three robot-view example frames from the real-world HRI corpus: (a) a single-participant interaction, (b) a two-participant interaction, and (c) a three-participant interaction.}
  \label{fig:robot_view_examples}
\end{teaserfigure}

\maketitle

\section{Introduction}
Social robots are increasingly being deployed in real-world environments such as commercial facilities and public spaces (e.g., \cite{Kanda2010ShoppingMall,Niemela2019ShoppingMall}). In such settings, spontaneous interactions arise in which passersby initiate engagement with a robot without prior preparation or instruction~\cite{Nielsen2023UnguidedInteractions}. In these interactions, users are free to leave at any time and the timing of conversation onset and termination is unconstrained. Furthermore, multiple participants may naturally join the conversation. This uncontrolled nature creates conditions fundamentally different from laboratory settings, where experimenters can regulate participant behavior and conversational flow, and introduces unique challenges for evaluating interaction quality~\cite{JungHinds2018RobotsWild}.

Evaluating and modeling interaction quality in such real-world settings is therefore an important research challenge. If interaction quality can be reliably estimated, the resulting estimates can inform improvements to dialogue strategies and, ultimately, enable robots to autonomously adapt their behavior. However, much prior work on interaction-quality modeling and evaluation has been developed and studied under controlled laboratory conditions, and it remains unclear whether such findings generalize to real-world deployments where user disengagement is unconstrained~\cite{JungHinds2018RobotsWild}.

As a first step toward addressing this challenge, this study focuses on rapport as a measure of interaction quality. Rapport is a concept reflecting the relational quality of an interaction, and the Connection-Coordination Rapport (CCR) Scale has recently been proposed as a rapport measure specific to human-robot interaction (HRI)~\cite{Lin2025CCRScale}. We construct a dataset of human-robot interactions collected in a real retail environment---specifically, a drugstore in Japan---where a remotely operated (Wizard of Oz) social robot was placed near the store entrance, allowing customers to freely initiate and terminate interactions. Third-party CCR annotations were applied to this dataset, and we establish a multimodal baseline model for automatic rapport estimation. To the best of our knowledge, this is one of the first studies to automatically estimate third-party-rated rapport from multimodal recordings of WoZ-mediated real-world HRI in an uncontrolled retail environment.

The contributions of this study are threefold:
\begin{itemize}
\item We construct one of the first real-world HRI datasets with rapport annotation, collected in an uncontrolled retail environment.
\item We establish a multimodal baseline for automatic rapport estimation in WoZ-mediated real-world HRI and show that fusing zero-shot large language model (LLM) predictions with embedding-based models yields the best performance on this dataset.
\item We present an empirical analysis of rapport in WoZ-mediated real-world HRI, offering insights into the characteristics of naturally occurring human-robot interactions, such as users' free entry and exit and multi-party participation, that distinguish them from controlled laboratory settings.
\end{itemize}

\section{Related Work}
\label{sec:related_work}
\subsection{Rapport for Interaction Quality in HRI}

Rapport refers to the quality of the relationship that emerges between interaction partners during an interaction. Tickle-Degnen and Rosenthal \cite{TickleDegnenRosenthal1990NatureRapport} conceptualized rapport as a dynamic structure consisting of three components: mutual attentiveness, positivity, and coordination. Rather than a stable individual personality trait, rapport is understood as a dyadic property that emerges through interaction.

Rapport is important because it is not merely an impression-based judgment, but is closely tied to successful relationship building and interaction. In interpersonal interaction research, rapport has been linked to outcomes such as improved learning in educational settings and successful negotiation (e.g., \cite{Nadler2004RapportNegotiation,SinhaCassell2015WeClick}). However, previous rapport scales have primarily relied on first-person evaluation \cite{Gratch2007CreatingRapport,Gratch2015Negotiation,NomuraKanda2016RERS}.

Lin et al. \cite{Lin2025CCRScale} proposed the Connection-Coordination Rapport (CCR) Scale for HRI, which enables rapport assessment from a third-person perspective. In HRI data, the CCR Scale exhibits a two-factor structure in which items related to mutual attentiveness and coordination cluster under the Coordination factor, whereas items related to interpersonal warmth, including positivity, cluster under the Connection factor. Furthermore, Lin et al. \cite{Lin2026ReducedCCRScale} proposed an eight-item reduced-length CCR Scale to reduce response burden and administration time and demonstrated its reliability and validity for third-party video-based assessment. In this study, we use the reduced-length CCR Scale, which is suitable for third-party annotation, to measure the relational quality of interaction in real-world HRI.

\raggedbottom
\subsection{Subjective Evaluation of Interaction Quality in HRI}

Evaluating interaction quality is a central challenge in HRI. In non-task-oriented human--agent and human--robot interactions, objective metrics such as task success or dialogue length are often insufficient, and subjective or relational criteria such as satisfaction, impression, enjoyment, and rapport have therefore been used to assess interaction quality~\cite{Wei2025AdversarialRapport,Pereira2024LLMEnjoyment,Santana2025SpeechToJoy,Wei2021Satisfaction,Wei2022DialogueExchangeImpression}.

Wei et al. modeled dialogue-level user satisfaction~\cite{Wei2021Satisfaction} and further showed that multitask learning can leverage the relationship between exchange-level sentiment and dialogue-level impression~\cite{Wei2022DialogueExchangeImpression}. In controlled human--agent interactions, Cerekovic et al.~\cite{Cerekovic2017RapportVirtualAgents} predicted both self-reported and observer-rated rapport from audiovisual social cues and personality measures. In HRI, Pereira et al.~\cite{Pereira2024LLMEnjoyment} examined LLM-based enjoyment detection, while Santana et al.~\cite{Santana2025SpeechToJoy} predicted self-reported enjoyment from pretrained audio and text embeddings. In a related multimodal dialogue setting, Wei et al.~\cite{Wei2025AdversarialRapport} investigated dialogue-level rapport recognition.

However, most of this work has been conducted in laboratory or online settings where interaction timing, participant roles, and conversational structure are relatively controlled. In real-world HRI, by contrast, interaction onset and termination, duration, and the number of participants are often not fixed, and public-space interactions are shaped by surrounding people and environmental contingencies~\cite{JungHinds2018RobotsWild,BenYoussef2017UEHRI,Nielsen2023UnguidedInteractions}. Field studies in retail and public environments show that social robots can be deployed in practice~\cite{Kanda2010ShoppingMall,Niemela2019ShoppingMall}. However, automatic estimation of interaction quality in real-world environments remains underexplored. Our study addresses this gap by estimating third-party CCR rapport scores from multimodal data collected in a real-world retail setting.

\section{Dataset with Rapport Annotation}
\label{sec:method}
\subsection{Real-World HRI Corpus}
In this study, we used an interaction dataset recorded at a drugstore (selling cosmetics, daily goods, and medicine) in Japan, where a teleoperated service robot provided store guidance and casual conversation to visiting customers. The robot was operated for a total of 32 hours over six days, resulting in 131 recorded sessions. The data were collected with approval from an institutional ethics committee.

The robot used in the HRI sessions was Sota (produced by Vstone)\footnote{\url{https://sota.vstone.co.jp/home/}}, a tabletop humanoid robot approximately 0.3 m tall. Sota is equipped with a built-in microphone, and a camera was mounted behind it, enabling the operator to monitor both audio and video in real time while controlling the robot's speech and gestures. The operator's voice was captured via a microphone, converted into a robot-like voice, and played through the robot's speaker, while gestures were controlled using a controller. The robot's gaze was controlled using an eye-tracking device. Further details regarding data collection are provided by Yano et al.~\cite{Yano2026Breaking}.

From the 131 recorded sessions, we applied a series of filters to retain sessions suitable for rapport annotation. Specifically, we excluded sessions involving four or more participants, as large groups frequently engaged in side conversations among themselves independent of the robot, making it difficult to isolate and reliably assess human-robot rapport. We also excluded sessions in which a preschool-aged child (under approximately 6 years of age) was present, as such participants rarely engaged in verbal communication with the robot and exhibited interaction styles markedly different from those of other visitors; we consider this population a subject for independent future investigation. Finally, sessions containing fewer than two user utterances across the entire session were excluded, as interactions of such limited verbal content were deemed insufficient for meaningful rapport assessment.

After applying these criteria, the dataset was reduced to 62 sessions comprising 101 participants in total. Excluding the four who produced no usable speech, the remaining 97 participants were included as analysis targets. On average, each analyzed participant produced $7.06$ ($SD = 5.83$) utterances. At the session level, each session contained an average of $11.85$ ($SD = 9.08$) utterances with a mean video duration of $54.23$ ($SD = 42.42$) seconds (see Table~\ref{tab:dataset_stats}).

\flushbottom
\subsection{Rapport Annotation}
\label{sec:rapport_annotation}
Third-party rapport annotations were assigned to the human-robot interactions described above. Rapport was measured using the eight-item Connection-Coordination Rapport Scale (CCR-8)~\cite{Lin2026ReducedCCRScale}, a short-form rapport scale for HRI whose reliability and validity have been demonstrated for third-party video-based assessment. The scale comprises two four-item factors: Connection and Coordination. Each item was rated on a five-point Likert scale ranging from 1 (strongly disagree) to 5 (strongly agree). The overall rapport score was computed as the mean of all eight items. As the CCR-8 was originally developed in English, the items were translated into Japanese through careful discussion.

Three third-party annotators watched the interaction videos and provided rapport ratings. All three annotators were Japanese speakers and had no conflict of interest with respect to our study; one was a professional service staff member, providing an evaluative perspective grounded in real-world customer interaction experience. Annotators were trained using the CCR-8 item definitions and example videos prior to the main annotation phase. In addition, to supplement the item definition of each of the eight items, example behavioral criteria adapted to the real-world service setting were established in advance through discussion between the researchers and the annotators and shared with all annotators.

Rapport was assessed at the individual level: each annotator rated the rapport that each individual participant appeared to hold toward the robot, regardless of whether the session involved a single participant or multiple participants. For the 34 multi-participant sessions, group-level ratings were additionally collected to capture the rapport of the group as a whole toward the robot; however, only the individual-level ratings are used in our experiments.

\begin{table}[t]
\caption{Dataset statistics after filtering.}
\label{tab:dataset_stats}
\small
\setlength{\tabcolsep}{4pt}
\begin{tabular*}{\columnwidth}{@{\extracolsep{\fill}}lr@{}}
\toprule
Statistic & \multicolumn{1}{c}{$M \pm SD$} \\
\midrule
\textit{Individual level} ($n = 97$) & \\
\quad \# Utterances & $7.06 \pm 5.83$ \\
\addlinespace
\textit{Session level} ($n = 62$) & \\
\quad Video duration (seconds) & $54.23 \pm 42.42$ \\
\quad \# Utterances & $11.85 \pm 9.08$ \\
\midrule
\textit{Session composition} & \\
\quad \# Individual sessions & 28 \\
\quad \# Multi-participant sessions & 34 \\
\midrule
\textit{Rapport scores (individual level, 3 annotators, $n = 97$)} & \\
\quad Score & $3.72 \pm 0.80$ \\
\quad ICC(2,3) & .85 \\
\quad Cronbach's $\alpha$ & .95 \\
\addlinespace
\textit{Rapport scores (group level, 3 annotators, $n = 34$)} & \\
\quad Score & $3.71 \pm 0.81$ \\
\quad ICC(2,3) & .86 \\
\quad Cronbach's $\alpha$ & .96 \\
\bottomrule
\end{tabular*}
\end{table}

\subsection{Annotation Statistics}
The final CCR score used in all analyses was the mean of the three annotators' ratings. Before using the aggregate CCR-8 score, we assessed the internal consistency of its items using Cronbach's $\alpha$. At the individual level, using item scores averaged across the three annotators, $\alpha$ was $.93$, $.92$, and $.95$ for the Connection factor, Coordination factor, and overall rapport score, respectively. These high values indicate strong internal consistency. Together with the scoring procedure specified in the original CCR-8 paper~\cite{Lin2026ReducedCCRScale}, they support the use of the aggregate rapport score in the present analyses.

Inter-rater reliability of the rapport score was then estimated using two-way absolute-agreement intraclass correlation coefficients (ICC)~\cite{KooLi2016ICC}. Because the analyses relied on averaged ratings, average-measures ICC(2,3) is reported as the primary reliability statistic. At the individual level, ICC(2,3) for the overall rapport score was $.85$; at the group level, it was $.86$. Both values indicate good agreement among the three annotators, justifying the use of their mean rating as a stable and reliable measure of rapport. All subsequent analyses therefore use the mean of the three annotators' individual-level CCR ratings as the rapport measure.

Descriptive statistics for the annotated rapport scores are summarized in Table~\ref{tab:dataset_stats}. The individual-level rapport score yielded $M = 3.72$ ($SD = 0.80$), and the group-level score yielded $M = 3.71$ ($SD = 0.81$).
Individual-level rapport scores ranged from 1.38 to 4.88.
Illustrative low-rapport interactions included cases in which a participant ignored the robot's prompts or walked away mid-utterance, whereas high-rapport interactions included participants who actively asked questions and sustained the exchange with laughter.

\section{Rapport Estimation Model}
\label{sec:rapport_model}
\subsection{Task Definition}
\label{sec:task_definition}
Let $\mathcal{I} = \{I_1, I_2, \dots, I_N\}$ denote a set of $N$ interactions. Each interaction $I_i$ may involve one or more participants. For each participant in an interaction, we create an individual instance consisting of their observed features and corresponding rapport score.

For each instance, we observe a participant's time series of feature vectors for a given modality $m \in \{\text{text},\, \text{audio},\, \text{visual}\}$:
\begin{equation}
    \mathbf{X}^{(m)} = \bigl(\mathbf{x}_{1}^{(m)},\, \mathbf{x}_{2}^{(m)},\,
    \dots,\, \mathbf{x}_{T}^{(m)}\bigr),
\end{equation}
where $T$ denotes the sequence length. Each embedding-based model receives a single-modality input $\mathbf{X}^{(m)}$.
The estimation target is a scalar rapport score $y \in \mathbb{R}$ for that participant. The goal of the rapport estimation task is to learn a function
\begin{equation}
    f_\theta:\; \mathbf{X}^{(m)} \;\longmapsto\; \hat{y},
\end{equation}
parameterized by $\theta$, such that $\hat{y}$ approximates $y$.

Because rapport scores are collected from human annotators, absolute score values inevitably contain annotator-specific noise arising from individual differences in rating style and scale usage. We therefore seek to capture the linear association between predicted and target scores while penalizing discrepancies in their means and variances. Accordingly, we adopt the Concordance Correlation Coefficient (CCC)~\cite{Lin1989CCC} as the basis for the training objective. CCC is defined as
\begin{equation}
    \rho_c = \frac{2\,\sigma_{\hat{y}y}}
                 {\sigma_{\hat{y}}^{2} + \sigma_{y}^{2}
                  + (\bar{\hat{y}} - \bar{y})^{2}},
\end{equation}
where $\sigma_{\hat{y}y}$ is the covariance between predictions and targets, $\sigma_{\hat{y}}^2$ and $\sigma_y^2$ are their respective variances, and $\bar{\hat{y}}$, $\bar{y}$ are their means. CCC jointly captures Pearson correlation (linear association) and mean/variance agreement (absolute calibration), with $\rho_c \in [-1, 1]$ and $\rho_c = 1$ indicating perfect agreement. The training loss is defined as
\begin{equation}
    \mathcal{L} = 1 - \rho_c,
\end{equation}
which is minimized when predictions are maximally concordant with the ground-truth rapport scores.

\subsection{Pretrained Embedding-Based Feature Extraction}
\subsubsection{Preprocessing}
We first prepared aligned multimodal data for each participant through automatic annotation followed by manual correction.

\paragraph{Participant Bounding Boxes}
Participant bounding boxes were detected using a DETR-based person detection model from DEIMv2-Wholebody34~\cite{DEIMv2-Wholebody34}. Bounding boxes were extracted for each frame to localize individual participants throughout the interaction.

\paragraph{Transcription and Speaker Identification}
Speech transcription was performed using Whisper-large~\cite{Radford2023Whisper}, which automatically segmented the audio and generated utterance-level transcripts with timestamps. Each utterance was then manually annotated with a participant ID, allowing us to link transcripts, audio segments, and video clips to individual participants.

While some steps involved manual annotation (e.g., speaker ID assignment), our focus in this work is on rapport estimation rather than fully automatic preprocessing. Developing end-to-end automatic methods for participant tracking and identification remains an important direction for future work.

\subsubsection{Feature Extraction}
We extract fixed representations for each participant using frozen pretrained models across three modalities: text, audio, and visual. Text and audio representations are extracted at the utterance level, taking as input the transcribed utterance text and the utterance-segmented speech waveform, respectively. Visual representations are extracted at the clip level.

\paragraph{Text}
Text representations were computed at the utterance level using Sentence-T5-large~\cite{Ni2022SentenceT5} (hereafter ST5). Each transcribed utterance text was encoded into a 768-dimensional embedding, which was then L2-normalized.

\paragraph{Audio}
Audio representations were computed at the utterance level using HuBERT-large-ll60k~\cite{Hsu2021HuBERT}. Speech segments were first isolated by cutting the recording at utterance boundaries derived from transcript timestamps. Each resulting waveform segment was encoded into a 1024-dimensional embedding, which was then L2-normalized.

\paragraph{Visual}
Visual representations were computed at the clip level using V-JEPA 2.1 ViT-Gigantic~\cite{MurLabadia2026VJEPA21} (hereafter V-JEPA). Videos were sampled at approximately 8\,fps and divided into non-overlapping 64-frame clips, producing a sequence of 1664-dimensional features from each clip. The features were then L2-normalized. When target-user bounding boxes were available, weighted spatial pooling was applied to focus on the target participant. The resulting temporal feature sequence was aggregated using the additive attention pooling described below.

\subsection{Embedding-Based Models}
Each modality stream shares the same two-stage architecture,
based on the model proposed by Santana et al.~\cite{Santana2025SpeechToJoy}:
an additive attention pooling module that aggregates the variable-length
feature sequence into a single interaction embedding,
followed by a prediction head that regresses the rapport score from
that embedding.
A separate model is trained for each modality (text, audio, and visual).
\subsubsection{Additive Attention Pooling}
Given the feature sequence
$\mathbf{X}_i^{(m)} = (\mathbf{x}_{i,1}^{(m)}, \dots, \mathbf{x}_{i,T_i}^{(m)})$
of length $T_i$ for interaction $I_i$,
a learned linear scoring function computes a scalar score for each step:
\begin{equation}
    s_{i,t} = \mathbf{w}^\top \mathbf{x}_{i,t}^{(m)} + b,
\end{equation}
where $\mathbf{w} \in \mathbb{R}^d$ and $b \in \mathbb{R}$ are
modality-specific learnable parameters.
Padding positions are masked before
the softmax, and the attention weights are computed as
\begin{equation}
    \alpha_{i,t} = \frac{\exp(s_{i,t})}{\sum_{t'=1}^{T_i} \exp(s_{i,t'})}.
\end{equation}
The interaction embedding is then obtained as the weighted sum
\begin{equation}
    \mathbf{v}_i = \sum_{t=1}^{T_i} \alpha_{i,t}\, \mathbf{x}_{i,t}^{(m)} \in \mathbb{R}^d.
\end{equation}
\subsubsection{Prediction Head}
The interaction embedding $\mathbf{v}_i$ is passed to a two-layer MLP:
\begin{equation}
    \hat{y}_i = \mathbf{W}_2\,\mathrm{Dropout}(\mathrm{ReLU}(\mathbf{W}_1 \mathbf{v}_i + \mathbf{b}_1)) + b_2,
\end{equation}
where Dropout with rate 0.2 is applied
after the activation.
The output $\hat{y}_i$ is a scalar predicted on the z-score scale,
which is inverse-transformed to the original rapport score scale
before evaluation.

\subsection{Zero-Shot LLMs}
We evaluated three LLMs as zero-shot prompt-based comparators:
GPT-5.4, Claude Sonnet 4.6,
and Gemini 2.5 Flash.
Each model was prompted to act as a third-party annotator and rate
the eight CCR-8 items on the same 5-point scale used by human annotators.
As with the human annotation procedure, the prompt (written in Japanese)
included, for each item, the item definition and example behavioral criteria
described in Section~\ref{sec:rapport_annotation}.
The model returned a structured JSON object, which was parsed into
item scores and averaged to obtain the predicted rapport score
for each participant.
Depending on the model's multimodal capabilities, we varied the input
modalities across three conditions: text-only~(T), text+audio~(T+A),
and text+audio+visual~(T+A+V).
In all conditions, the transcript with speaker-ID prefixes was provided
as the text input.
In the T+A condition, the corresponding wav file was additionally supplied.
In the T+A+V condition, the video with per-speaker bounding boxes
was further included to provide visual context.
Gemini models were queried using the API-default thinking configuration; Claude was queried without extended thinking; GPT-5.4 was queried with reasoning effort \textit{none}. All LLM predictions were obtained through the respective public APIs in April 2026, with a single run per condition.

\subsection{Late Fusion}
To combine predictions from multiple models, we use unweighted averaging of the prediction scores. We also examined weighted averaging, but its performance did not differ substantially from unweighted averaging; we therefore adopted the equal-weight scheme, which requires no additional parameters.

\subsection{Evaluation Procedure}
\subsubsection{Data Splits}
The embedding-based models and the Random Baseline were evaluated under
the same predefined 30-fold train/val/test split.
The splits were constructed at the session level,
such that all participants from the same session are assigned to
the same set, ensuring no data leakage across splits.
In each fold, the held-out test set was excluded from all stages
of training, including model selection and early stopping.
A validation set sampled from the remaining non-test data was used
for the same purposes.
Final metrics were computed globally by pooling the out-of-fold test
predictions across all folds, rather than averaging per-fold metrics.
For the Random Baseline, test predictions in each fold were sampled with
replacement from the target-score distribution of the corresponding training
partition; this procedure was repeated with 100 random seeds, and the table
reports the mean metrics across seeds.

\subsubsection{Metrics}
We report three complementary regression metrics.
Mean Absolute Error (MAE) measures the average magnitude of prediction
errors in the original score scale.
Pearson correlation coefficient (PCC) captures the linear association between
predicted and ground-truth scores.
Concordance Correlation Coefficient
(CCC; see Section~\ref{sec:task_definition}) jointly accounts for both correlation and
mean/variance agreement, and serves as the primary metric consistent
with the training objective.
Lower MAE and higher PCC/CCC indicate better performance.

\paragraph{Use of AI Tools}
Parts of the experiment, analysis, and figure-generation code used in this
study were developed with the assistance of generative AI coding tools
(Claude, Anthropic; Codex, OpenAI).

\section{Experimental Results}
\label{sec:results}

\subsection{Rapport Estimation Performance}
\label{sec:rapport_prediction}

Results are summarized in Table~\ref{tab:prediction_no_closing}.

\setlength{\textfloatsep}{6pt plus 1pt minus 1pt}
\begin{table}[!h]
\centering
\caption{Comparison of CCR rapport estimation performance in real-world HRI. The best value for each metric is shown in bold with underline.}
\label{tab:prediction_no_closing}
\footnotesize
\setlength{\tabcolsep}{2pt}
\renewcommand{\arraystretch}{0.92}
\begin{tabular*}{\columnwidth}{@{}p{0.36\columnwidth}@{\extracolsep{\fill}}lccc@{}}
\toprule
Model & Modality & MAE $\downarrow$ & PCC $\uparrow$ & CCC $\uparrow$ \\
\midrule
\multicolumn{5}{l}{\textit{Baseline}} \\
\quad Random Baseline & -- & 0.901 & $-0.004$ & $-0.005$ \\
\midrule
\multicolumn{5}{l}{\textit{Zero-shot LLMs}} \\
\quad GPT-5.4          & T     & 0.617 & $0.644$                           & 0.555             \\
\quad Claude Sonnet 4.6 & T & 0.741 & $0.619$ & 0.477 \\
\quad Gemini 2.5 Flash & T     & 0.634 & $\textbf{\underline{0.665}}$      & 0.580             \\
\quad Gemini 2.5 Flash & T+A   & 0.592 & $0.602$            & 0.550             \\
\quad Gemini 2.5 Flash & T+A+V & 0.549 & $0.625$                           & \textbf{\underline{0.618}} \\
\midrule
\multicolumn{5}{l}{\textit{Pretrained Embeddings}} \\
\quad ST5              & T     & 0.633 & $0.327$ & 0.281 \\
\quad HuBERT           & A     & 0.616 & $0.464$ & 0.460 \\
\quad V-JEPA       & V     & 0.666 & $0.331$ & 0.310 \\
\midrule
\multicolumn{5}{l}{\textit{Late Fusion (unweighted average)}} \\
\quad ST5+HuBERT          & T+A   & 0.565             & $0.514$ & 0.444 \\
\quad ST5+V-JEPA          & T+V   & 0.603             & $0.443$ & 0.341 \\
\quad HuBERT+V-JEPA       & A+V   & 0.542                           & $0.526$ & 0.465 \\
\quad ST5+HuBERT+V-JEPA   & T+A+V & $\textbf{\underline{0.540}}$   & $0.567$ & 0.444 \\
\bottomrule
\end{tabular*}
\vspace{-0.6em}
\end{table}

Zero-shot LLMs demonstrated strong performance in real-world rapport estimation. In terms of CCC, Gemini 2.5 Flash (T+A+V) achieved the highest score in the table (0.618), while, notably, Gemini 2.5 Flash (T) attained the highest PCC (0.665). Furthermore, the second-best PCC result, GPT-5.4 (T) = 0.644, also relied solely on text input, suggesting that the text modality alone carries substantial useful information for rapport estimation.

For Gemini 2.5 Flash (T), the subscale CCCs were 0.614 for Connection and 0.473 for Coordination.

Embedding-based models exhibited considerable variation in performance across modalities. The text encoder ST5 (T) yielded the lowest PCC and CCC among all supervised models, highlighting the difficulty of rapport estimation from text embeddings alone. In contrast, the audio encoder HuBERT (A) outperformed all other supervised single-modality models across all three metrics.

When modalities were integrated via late fusion, none of the combinations surpassed the zero-shot LLMs in terms of PCC or CCC. Nevertheless, the fused models tended to achieve lower MAE values, with ST5+HuBERT+V-JEPA (T+A+V) attaining the lowest MAE in the table.

\subsection{Complementarity Analysis}
\label{sec:complementarity}

\begin{figure}[t]
\centering
\includegraphics[width=0.92\columnwidth]{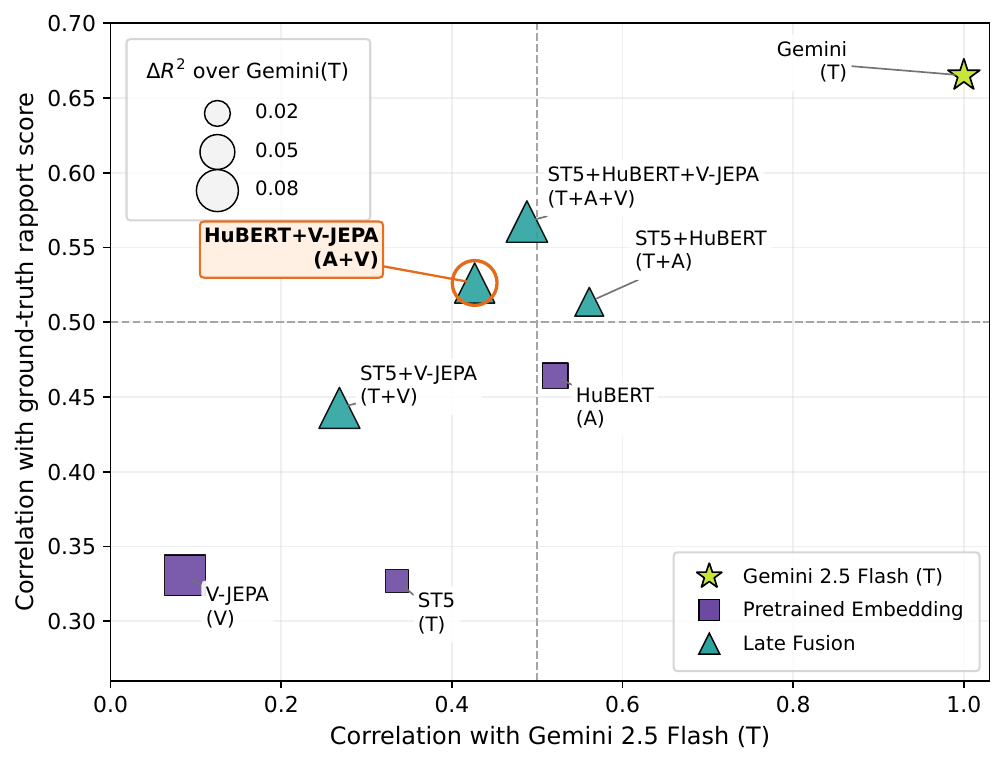}
\caption{Utility--redundancy map relative to Gemini 2.5 Flash (T). Point size indicates incremental explanatory value over Gemini (T).}
\label{fig:prediction_complementarity_map}
\Description{Scatter plot of model utility versus redundancy relative to Gemini 2.5 Flash (T). The vertical axis is correlation with the ground-truth rapport score, the horizontal axis is correlation with Gemini 2.5 Flash (T) predictions, and point size reflects incremental R-squared over Gemini (T). HuBERT+V-JEPA (A+V) is highlighted as a complementary candidate.}
\end{figure}

The results in Section~\ref{sec:rapport_prediction} show that multimodal LLMs such as Gemini 2.5 Flash (T+A+V) achieved the highest CCC, and Gemini 2.5 Flash (T) delivered competitive performance across all metrics using text input alone. However, multimodal LLMs that directly process audio and video may introduce greater inference latency and monetary cost than text-only LLMs. We therefore focus on Gemini 2.5 Flash (T) as the primary predictor and investigate whether embedding-based audio and visual models can complement it. Results are visualized in Figure~\ref{fig:prediction_complementarity_map}.

Figure~\ref{fig:prediction_complementarity_map} shows a utility--redundancy map in which the vertical axis represents each model's correlation with the ground-truth rapport score (standalone utility), the horizontal axis represents its correlation with Gemini (T) predictions (redundancy), and marker size encodes the incremental explanatory power $\Delta R^2$ when added to Gemini (T). Models in the upper-left region are therefore both informative and non-redundant with respect to Gemini (T), making them natural candidates for fusion.

Among the embedding-based models, ST5+HuBERT+V-JEPA (T+A+V) and HuBERT+V-JEPA (A+V) occupied the upper-left region of the map. Because the fusion analysis replaces the text branch with Gemini (T), the natural complementary candidates are the audio and visual streams that Gemini (T) cannot access; HuBERT+V-JEPA (A+V) combines both non-text modalities while occupying this informative and non-redundant region. We therefore focus on HuBERT+V-JEPA (A+V) as the complementary model. While we focus on A+V, for completeness we also evaluate fusion with each of HuBERT and V-JEPA individually.

To quantify complementarity, we employ two metrics. The partial correlation $r(y, \hat{y}^{AV} \mid \hat{y}^{G})$ measures the association between HuBERT+V-JEPA (A+V) predictions and the ground-truth rapport score after partialling out the variance explained by Gemini (T). The incremental $R^2$ is defined as
\begin{equation}
  \Delta R^2 = R^2(y \sim \hat{y}^{G} + \hat{y}^{AV}) - R^2(y \sim \hat{y}^{G}),
\end{equation}
where $\hat{y}^{G}$ and $\hat{y}^{AV}$ denote the predictions of Gemini (T) and HuBERT+V-JEPA (A+V), respectively. For HuBERT+V-JEPA (A+V), we obtained $r(y, \hat{y}^{AV} \mid \hat{y}^{G}) = 0.359$ and $\Delta R^2 = 0.072$. These results indicate that, although HuBERT+V-JEPA (A+V) does not match Gemini (T) in standalone performance, it captures rapport-relevant signals that Gemini (T) alone does not account for, and thus has the potential to improve estimation quality when combined via fusion.

\subsection{Fusion of LLM with Embedding-Based Models}
\label{sec:closing_aware_llm_fusion}

Table~\ref{tab:prediction_closing_fusion} presents the results of fusing Gemini 2.5 Flash (T) with embedding-based models via unweighted prediction-level averaging: each configuration averages the Gemini (T) prediction with one complementary embedding-based predictor. For the T+A+V configuration, the complementary predictor is the equal-weight average of the HuBERT and V-JEPA predictions (A+V) analyzed in Section~\ref{sec:complementarity}, so the effective weights are $1/2$ for Gemini (T) and $1/4$ each for HuBERT and V-JEPA. All three fusion configurations performed better than the reference baselines across all three metrics.

\begin{table}[!htb]
\centering
\caption{Unweighted prediction-level fusion of Gemini 2.5 Flash (T) with embedding-based models. The best value for each metric is shown in bold with underline.}
\label{tab:prediction_closing_fusion}
\small
\setlength{\tabcolsep}{3pt}
\renewcommand{\arraystretch}{0.92}
\begin{tabular*}{\columnwidth}{@{}p{0.40\columnwidth}@{\extracolsep{\fill}}lccc@{}}
\toprule
Model & Modality & MAE $\downarrow$ & PCC $\uparrow$ & CCC $\uparrow$ \\
\midrule
\multicolumn{5}{l}{\textit{Reference baselines (from Table~\ref{tab:prediction_no_closing})}} \\
Gemini 2.5 Flash          & T+A+V & 0.549 & $0.625$ & 0.618 \\
ST5+HuBERT+V-JEPA         & T+A+V & 0.540 & $0.567$ & 0.444 \\
\midrule
\multicolumn{5}{l}{\textit{Unweighted-average fusion with Gemini 2.5 Flash (T)}} \\
+ HuBERT          & T+A   & 0.507 & $0.660$ & 0.629 \\
+ V-JEPA          & T+V   & 0.490 & $0.712$ & 0.632 \\
+ HuBERT+V-JEPA   & T+A+V & \textbf{\underline{0.471}} & $\textbf{\underline{0.717}}$ & \textbf{\underline{0.656}} \\
\bottomrule
\end{tabular*}
\end{table}

The best-performing configuration, Gemini (T) + HuBERT+V-JEPA (T+A+V), achieved an MAE of 0.471, a PCC of 0.717, and a CCC of 0.656.

\begin{figure*}[!t]
\centering
\includegraphics[width=0.88\textwidth]{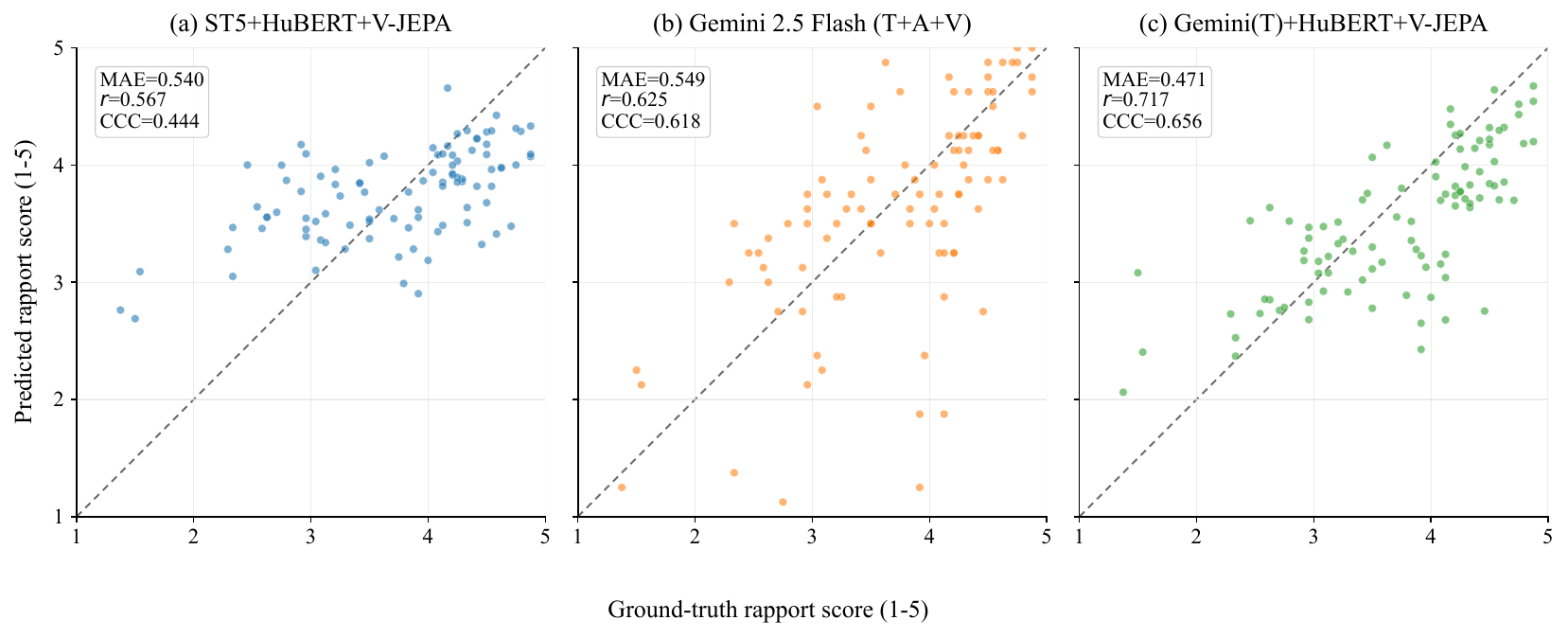}
\caption{Predicted versus ground-truth CCR rapport scores for three representative prediction conditions. Dashed lines indicate perfect prediction.}
\label{fig:closing_aware_prediction_scatter}
\Description{Three scatter plots of predicted versus ground-truth CCR rapport scores for the embedding-only T+A+V fusion, the strongest zero-shot LLM, and the unweighted-average fusion of Gemini text with HuBERT and V-JEPA predictions. The Gemini-plus-ML fusion panel shows the highest CCC among the three.}
\end{figure*}

\raggedbottom
\section{Discussion}
\label{sec:discussion}
This study demonstrates that rapport scores in real-world HRI can be estimated to a meaningful degree using both zero-shot LLMs and embedding-based models. In particular, a text-only LLM proved effective as a standalone predictor, and fusing it with audio-visual embedding-based models yielded the best overall performance.

\subsection{Comparison with Laboratory-Based HRI Studies}
\label{sec:comparison_lab}

We primarily compare our results with Speech-to-Joy \cite{Santana2025SpeechToJoy}, a laboratory-based HRI study that served as a methodological reference for our experimental design.
However, direct quantitative comparison should be interpreted with caution, as that work targets enjoyment rather than rapport and uses self-reported labels rather than the third-party annotations used in our study.

In our real-world setting, the LLM-based predictor outperforms the embedding-based late fusion model, and text embeddings alone prove weak, offering little useful signal for estimation. This stands in contrast to the findings of \cite{Santana2025SpeechToJoy}, where the late fusion of text and audio embedding-based models (T+A) outperforms LLM-based approaches on correlation metrics (PCC and CCC), and text features prove nearly as informative as acoustic ones.
The substantially shorter interaction durations in our deployment---our analyzed interaction units average 53.6 seconds, compared with approximately seven minutes in Speech-to-Joy---constitute one possible explanation for this difference, as shorter interactions may provide less rapport-related information for embedding representations.
The complex dynamics of multi-party interactions may further hinder the performance of embedding-based models.
In contrast, the LLM-based predictor showed strong performance under these challenging conditions.

\subsection{Performance across Interaction-Duration Conditions}
\label{sec:interaction_duration}

A defining characteristic of real-world HRI deployments, in contrast to laboratory settings, is that the timing and duration of interactions cannot be controlled in advance.
Participants approach, engage, and disengage from the robot at their own discretion, resulting in substantial variability in interaction length.
Among the 97 individual-level interaction units analyzed in this study, interaction duration averaged 53.6 seconds ($SD = 41.1$) with a median of 40.0 seconds, ranging from as brief as 12 seconds to as long as 227 seconds.
To examine estimation performance across interaction-duration conditions, we partition the interactions into \textit{short} (${\leq}40$ seconds) and \textit{long} (${>}40$ seconds) conditions using the median as a threshold, and compare model performance across the two conditions.

Gemini 2.5 Flash (T) achieved a CCC of 0.563 in the short condition and 0.551 in the long condition, with a difference of 0.012 between the conditions.
The text-only supervised predictor ST5 (T) increased from 0.168 to 0.411, a difference of 0.243.
Among the unimodal audio and visual predictors, HuBERT (A) declined from 0.463 to 0.420, a decrease of 0.043, and V-JEPA (V) declined from 0.330 to 0.295, a decrease of 0.035.

Only the supervised text model ST5 (T) showed a pronounced performance difference between the two conditions, with a marked drop in the short condition. Gemini 2.5 Flash (T) exhibited the smallest variation across conditions, although HuBERT (A) and V-JEPA (V) also varied little.

\flushbottom
\subsection{Performance across Group-Size Conditions}
\label{sec:multiparty}

A second factor that distinguishes real-world HRI from laboratory studies is the prevalence of multi-party interactions. Whereas laboratory studies can enforce strictly dyadic interactions, real-world deployments cannot constrain who chooses to join an ongoing interaction with the robot. Our dataset reflects this reality: of the 97 individual-level samples analyzed, 69 involved participants engaged in multi-party interactions. Specifically, 28 samples came from one-person interactions, 56 from two-person interactions, and 13 from three-person interactions. Group size was determined from each session's full set of participants, so members who are not among the 97 estimation targets---for example, those who produced little or no usable speech---still count toward the group size of their session. Mean ground-truth rapport was similar in the one- and two-person conditions but lower in the three-person condition (3.774, 3.775, and 3.391, respectively). Given the small three-person subset ($n=13$), these descriptive differences should be interpreted cautiously.

\begin{figure}[t]
\centering
\includegraphics[width=0.96\columnwidth]{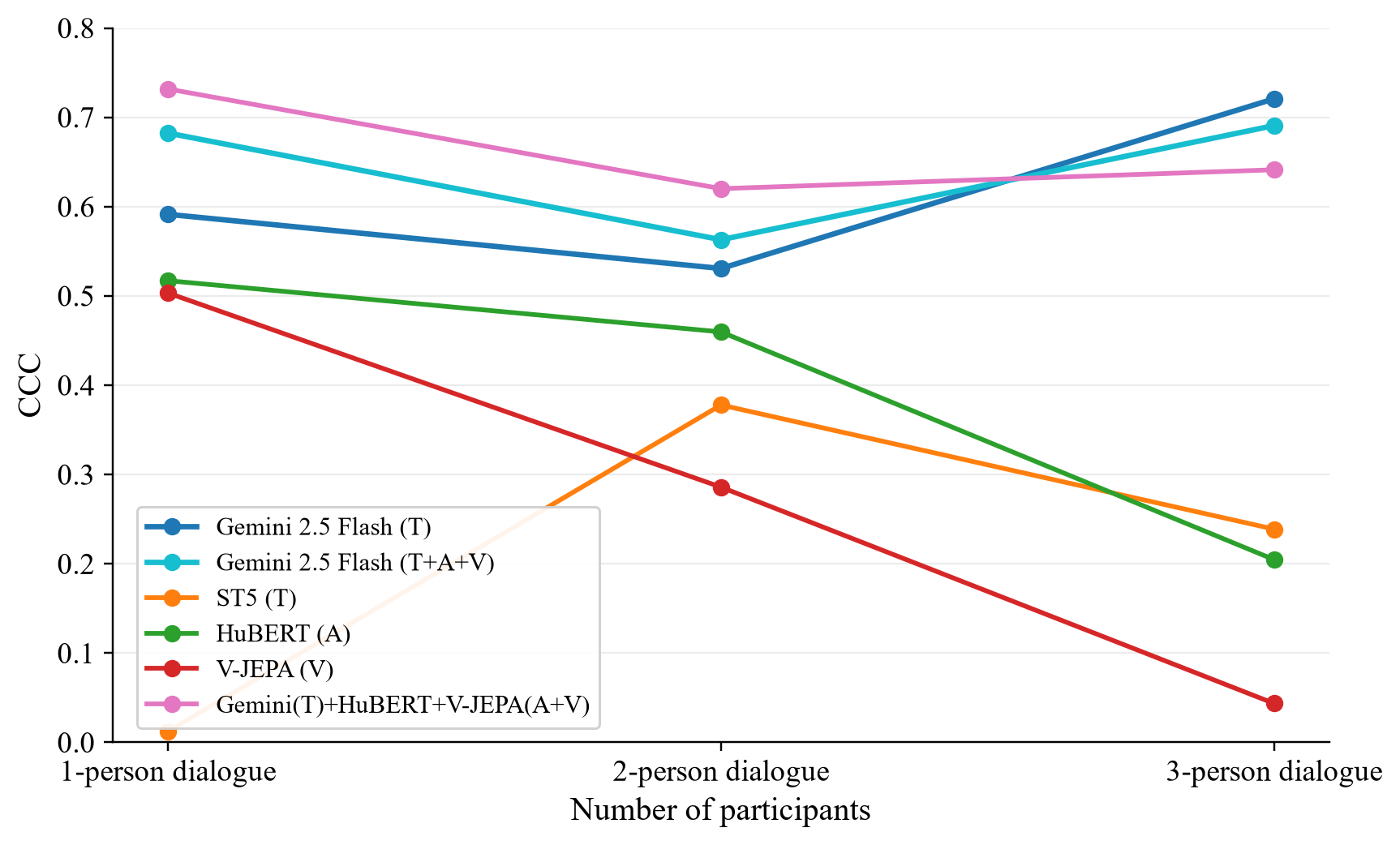}
\caption{CCC trends across one-, two-, and three-participant interactions for each prediction model.}
\label{fig:party_size_ccc_trends}
\Description{Line plot showing CCC values for each prediction model across one-, two-, and three-person interactions. Gemini 2.5 Flash (T) shows relatively high CCC values across the analyzed group-size conditions, whereas V-JEPA (V) declines as the number of participants increases.}
\end{figure}

As shown in Figure~\ref{fig:party_size_ccc_trends}, Gemini 2.5 Flash (T) consistently maintained high CCC across all group-size conditions. Interestingly, it achieved its highest performance in the three-person condition (CCC = 0.721), suggesting that richer multi-party discourse may actually provide the LLM with additional contextual cues that support rapport inference. In contrast, the supervised unimodal predictors---with the exception of ST5 (T)---exhibited a consistent decline as group size increased. In particular, the CCC of V-JEPA (V) decreased from 0.503 to 0.286 and then to 0.043.

Together with the findings in Section~\ref{sec:interaction_duration}, these results indicate that the supervised embedding-based predictors showed greater performance variation than Gemini 2.5 Flash (T) in our data. The variation was particularly large across group-size conditions. Whether this pattern reflects differences in the information available to the models or the limited training data available to the supervised models (Section~\ref{sec:limitations}) remains to be tested with larger datasets and additional LLMs.

\subsection{Limitations and Future Work}
\label{sec:limitations}

This study has several limitations. First, the data were collected in a single real-world service setting---a drugstore in Japan---and the participants were mainly Japanese speakers within a Japanese cultural context. It therefore remains unclear whether similar patterns would appear in other countries, languages, cultures, settings, or robot embodiments. In addition, because the robot's speech, gestures, and gaze were controlled by a human operator (WoZ), the rapport ratings may partly reflect the operator's conversational and social abilities. Accordingly, our findings concern WoZ-mediated social robot interaction, and generalizing them to autonomous HRI requires caution.

Second, the estimation experiments were based on 97 individual cases. Although real-world HRI data are valuable, this sample remains small for robust machine-learning evaluation, and the additional analyses by interaction duration and number of participants relied on even smaller subsets. These findings should therefore be interpreted as exploratory. Moreover, the 97 targets were limited to cases with available audio and transcripts, meaning that encounters with little or no speech or very early disengagement were underrepresented. The results therefore reflect only interactions in which some degree of engagement occurred and the required data were available.

Third, the target variable was not participants' self-reported rapport, but rapport judged by third parties from observable behavior and interaction content. Our findings therefore demonstrate the potential of third-party rapport estimation rather than direct estimation of participants' own experienced rapport, enjoyment, or satisfaction. In addition, although the Japanese translation of the CCR-8 showed good inter-rater agreement and high internal consistency, we did not conduct a full psychometric validation of the Japanese version.

Fourth, the zero-shot LLMs and embedding-based predictors were not evaluated under strictly identical conditions because they differed in input format and inference framework. The comparison should therefore be treated as a practical baseline, not evidence that LLMs are inherently superior. Zero-shot LLM performance may also vary with prompts, model versions, API specifications, and input formats. Finally, although Gemini 2.5 Flash performed strongly, the linguistic and interactional cues underlying its rapport estimates remain unclear and should be examined in future work.

\section{Conclusion}
\label{sec:conclusion}

In this study, we examined the automatic estimation of third-party-rated rapport scores in real-world HRI. The zero-shot LLM (Gemini 2.5 Flash (T)) achieved strong predictive performance from text-only input. Fusing Gemini (T) with HuBERT and V-JEPA yielded the best overall performance, suggesting that LLMs and embedding-based predictors are complementary rather than competing components. Future work should develop models robust to variation in interaction duration and multi-party interaction.

\section{Safe and Responsible Innovation Statement}
\label{sec:safe_responsible_innovation}

This study uses multimodal data from real-world HRI to estimate third-party-rated rapport and therefore requires careful attention to privacy protection and responsible use. Such estimation does not directly measure an individual's internal state and should not be used as the sole basis for evaluating or making decisions about individuals. Any deployment should restrict access to raw data, minimize retention, support de-identification, and provide clear notice and appropriate consent procedures.

\begin{acks}
We thank Atsuko Harita for her contributions to data annotation.
This work was partially supported by JSPS KAKENHI (No. 26K03016) and JST CREST (No. JPMJCR2563).
\end{acks}

\bibliographystyle{ACM-Reference-Format}
\bibliography{references}


\begin{thebibliography}{27}


\ifx \showCODEN    \undefined \def \showCODEN     #1{\unskip}     \fi
\ifx \showDOI      \undefined \def \showDOI       #1{#1}\fi
\ifx \showISBNx    \undefined \def \showISBNx     #1{\unskip}     \fi
\ifx \showISBNxiii \undefined \def \showISBNxiii  #1{\unskip}     \fi
\ifx \showISSN     \undefined \def \showISSN      #1{\unskip}     \fi
\ifx \showLCCN     \undefined \def \showLCCN      #1{\unskip}     \fi
\ifx \shownote     \undefined \def \shownote      #1{#1}          \fi
\ifx \showarticletitle \undefined \def \showarticletitle #1{#1}   \fi
\ifx \showURL      \undefined \def \showURL       {\relax}        \fi
\providecommand\bibfield[2]{#2}
\providecommand\bibinfo[2]{#2}
\providecommand\natexlab[1]{#1}
\providecommand\showeprint[2][]{arXiv:#2}

\bibitem[Ben-Youssef et~al\mbox{.}(2017)]%
        {BenYoussef2017UEHRI}
\bibfield{author}{\bibinfo{person}{Atef Ben-Youssef},
  \bibinfo{person}{Chlo{\'e} Clavel}, \bibinfo{person}{Slim Essid},
  \bibinfo{person}{Miriam Bilac}, \bibinfo{person}{Marine Chamoux}, {and}
  \bibinfo{person}{Angelica Lim}.} \bibinfo{year}{2017}\natexlab{}.
\newblock \showarticletitle{{UE-HRI}: A New Dataset for the Study of User
  Engagement in Spontaneous Human-Robot Interactions}. In
  \bibinfo{booktitle}{\emph{Proceedings of the 19th ACM International
  Conference on Multimodal Interaction}} \emph{(\bibinfo{series}{ICMI '17})}.
  \bibinfo{pages}{464--472}.
\newblock
\urldef\tempurl%
\url{https://doi.org/10.1145/3136755.3136814}
\showDOI{\tempurl}


\bibitem[Cerekovic et~al\mbox{.}(2017)]%
        {Cerekovic2017RapportVirtualAgents}
\bibfield{author}{\bibinfo{person}{Aleksandra Cerekovic}, \bibinfo{person}{Oya
  Aran}, {and} \bibinfo{person}{Daniel Gatica-Perez}.}
  \bibinfo{year}{2017}\natexlab{}.
\newblock \showarticletitle{Rapport with Virtual Agents: What Do Human Social
  Cues and Personality Explain?}
\newblock \bibinfo{journal}{\emph{IEEE Transactions on Affective Computing}}
  \bibinfo{volume}{8}, \bibinfo{number}{3} (\bibinfo{year}{2017}),
  \bibinfo{pages}{382--395}.
\newblock
\urldef\tempurl%
\url{https://doi.org/10.1109/TAFFC.2016.2545650}
\showDOI{\tempurl}


\bibitem[Gratch et~al\mbox{.}(2015)]%
        {Gratch2015Negotiation}
\bibfield{author}{\bibinfo{person}{Jonathan Gratch}, \bibinfo{person}{David
  DeVault}, \bibinfo{person}{Gale~M. Lucas}, {and} \bibinfo{person}{Stacy
  Marsella}.} \bibinfo{year}{2015}\natexlab{}.
\newblock \showarticletitle{Negotiation as a Challenge Problem for Virtual
  Humans}. In \bibinfo{booktitle}{\emph{Intelligent Virtual Agents}}
  \emph{(\bibinfo{series}{Lecture Notes in Computer Science},
  Vol.~\bibinfo{volume}{9238})}. \bibinfo{publisher}{Springer International
  Publishing}, \bibinfo{address}{Cham}, \bibinfo{pages}{201--215}.
\newblock
\urldef\tempurl%
\url{https://doi.org/10.1007/978-3-319-21996-7_21}
\showDOI{\tempurl}


\bibitem[Gratch et~al\mbox{.}(2007)]%
        {Gratch2007CreatingRapport}
\bibfield{author}{\bibinfo{person}{Jonathan Gratch}, \bibinfo{person}{Ning
  Wang}, \bibinfo{person}{Jillian Gerten}, \bibinfo{person}{Edward Fast}, {and}
  \bibinfo{person}{Robin Duffy}.} \bibinfo{year}{2007}\natexlab{}.
\newblock \showarticletitle{Creating Rapport with Virtual Agents}. In
  \bibinfo{booktitle}{\emph{Intelligent Virtual Agents}}
  \emph{(\bibinfo{series}{Lecture Notes in Computer Science},
  Vol.~\bibinfo{volume}{4722})}. \bibinfo{publisher}{Springer Berlin
  Heidelberg}, \bibinfo{address}{Berlin, Heidelberg},
  \bibinfo{pages}{125--138}.
\newblock
\urldef\tempurl%
\url{https://doi.org/10.1007/978-3-540-74997-4_12}
\showDOI{\tempurl}


\bibitem[Hsu et~al\mbox{.}(2021)]%
        {Hsu2021HuBERT}
\bibfield{author}{\bibinfo{person}{Wei-Ning Hsu}, \bibinfo{person}{Benjamin
  Bolte}, \bibinfo{person}{Yao-Hung~Hubert Tsai}, \bibinfo{person}{Kushal
  Lakhotia}, \bibinfo{person}{Ruslan Salakhutdinov}, {and}
  \bibinfo{person}{Abdelrahman Mohamed}.} \bibinfo{year}{2021}\natexlab{}.
\newblock \showarticletitle{{HuBERT}: Self-Supervised Speech Representation
  Learning by Masked Prediction of Hidden Units}.
\newblock \bibinfo{journal}{\emph{IEEE/ACM Transactions on Audio, Speech, and
  Language Processing}}  \bibinfo{volume}{29} (\bibinfo{year}{2021}),
  \bibinfo{pages}{3451--3460}.
\newblock
\urldef\tempurl%
\url{https://doi.org/10.1109/TASLP.2021.3122291}
\showDOI{\tempurl}


\bibitem[Hyodo(2025)]%
        {DEIMv2-Wholebody34}
\bibfield{author}{\bibinfo{person}{Katsuya Hyodo}.}
  \bibinfo{year}{2025}\natexlab{}.
\newblock \bibinfo{booktitle}{\emph{{DEIMv2-Wholebody34}: Lightweight Human
  Detection Models Generated on High-Quality Human Data Sets}}.
\newblock
\urldef\tempurl%
\url{https://github.com/PINTO0309/PINTO_model_zoo/tree/598e8d5d928e1d5a54511f34f6abede378831c2c/472_DEIMv2-Wholebody34}
\showURL{%
\tempurl}
\newblock
\shownote{Git commit 598e8d5d928e1d5a54511f34f6abede378831c2c; accessed
  2026-07-13}.


\bibitem[Jung and Hinds(2018)]%
        {JungHinds2018RobotsWild}
\bibfield{author}{\bibinfo{person}{Malte Jung} {and} \bibinfo{person}{Pamela
  Hinds}.} \bibinfo{year}{2018}\natexlab{}.
\newblock \showarticletitle{Robots in the Wild: A Time for More Robust Theories
  of Human-Robot Interaction}.
\newblock \bibinfo{journal}{\emph{ACM Transactions on Human-Robot Interaction}}
  \bibinfo{volume}{7}, \bibinfo{number}{1}, Article \bibinfo{articleno}{2}
  (\bibinfo{date}{May} \bibinfo{year}{2018}), \bibinfo{numpages}{5}~pages.
\newblock
\urldef\tempurl%
\url{https://doi.org/10.1145/3208975}
\showDOI{\tempurl}


\bibitem[Kanda et~al\mbox{.}(2010)]%
        {Kanda2010ShoppingMall}
\bibfield{author}{\bibinfo{person}{Takayuki Kanda}, \bibinfo{person}{Masahiro
  Shiomi}, \bibinfo{person}{Zenta Miyashita}, \bibinfo{person}{Hiroshi
  Ishiguro}, {and} \bibinfo{person}{Norihiro Hagita}.}
  \bibinfo{year}{2010}\natexlab{}.
\newblock \showarticletitle{A Communication Robot in a Shopping Mall}.
\newblock \bibinfo{journal}{\emph{IEEE Transactions on Robotics}}
  \bibinfo{volume}{26}, \bibinfo{number}{5} (\bibinfo{year}{2010}),
  \bibinfo{pages}{897--913}.
\newblock
\urldef\tempurl%
\url{https://doi.org/10.1109/TRO.2010.2062550}
\showDOI{\tempurl}


\bibitem[Koo and Li(2016)]%
        {KooLi2016ICC}
\bibfield{author}{\bibinfo{person}{Terry~K. Koo} {and} \bibinfo{person}{Mae~Y.
  Li}.} \bibinfo{year}{2016}\natexlab{}.
\newblock \showarticletitle{A Guideline of Selecting and Reporting Intraclass
  Correlation Coefficients for Reliability Research}.
\newblock \bibinfo{journal}{\emph{Journal of Chiropractic Medicine}}
  \bibinfo{volume}{15}, \bibinfo{number}{2} (\bibinfo{year}{2016}),
  \bibinfo{pages}{155--163}.
\newblock
\urldef\tempurl%
\url{https://doi.org/10.1016/j.jcm.2016.02.012}
\showDOI{\tempurl}


\bibitem[Lin(1989)]%
        {Lin1989CCC}
\bibfield{author}{\bibinfo{person}{Lawrence I-Kuei Lin}.}
  \bibinfo{year}{1989}\natexlab{}.
\newblock \showarticletitle{A Concordance Correlation Coefficient to Evaluate
  Reproducibility}.
\newblock \bibinfo{journal}{\emph{Biometrics}} \bibinfo{volume}{45},
  \bibinfo{number}{1} (\bibinfo{year}{1989}), \bibinfo{pages}{255--268}.
\newblock
\urldef\tempurl%
\url{https://doi.org/10.2307/2532051}
\showDOI{\tempurl}


\bibitem[Lin et~al\mbox{.}(2026)]%
        {Lin2026ReducedCCRScale}
\bibfield{author}{\bibinfo{person}{Ting-Han Lin}, \bibinfo{person}{Guan Chen},
  \bibinfo{person}{Bilge Mutlu}, \bibinfo{person}{J.~Gregory Trafton}, {and}
  \bibinfo{person}{Sarah Sebo}.} \bibinfo{year}{2026}\natexlab{}.
\newblock \showarticletitle{The Reduced-Length Connection-Coordination Rapport
  ({CCR}) Scale}.
\newblock \bibinfo{journal}{\emph{ACM Transactions on Human-Robot Interaction}}
  \bibinfo{volume}{15}, \bibinfo{number}{3}, Article \bibinfo{articleno}{57}
  (\bibinfo{year}{2026}), \bibinfo{numpages}{29}~pages.
\newblock
\urldef\tempurl%
\url{https://doi.org/10.1145/3796533}
\showDOI{\tempurl}


\bibitem[Lin et~al\mbox{.}(2025)]%
        {Lin2025CCRScale}
\bibfield{author}{\bibinfo{person}{Ting-Han Lin}, \bibinfo{person}{Hannah
  Dinner}, \bibinfo{person}{Tsz~Long Leung}, \bibinfo{person}{Bilge Mutlu},
  \bibinfo{person}{J.~Gregory Trafton}, {and} \bibinfo{person}{Sarah Sebo}.}
  \bibinfo{year}{2025}\natexlab{}.
\newblock \showarticletitle{Connection-Coordination Rapport ({CCR}) Scale: A
  Dual-Factor Scale to Measure Human-Robot Rapport}. In
  \bibinfo{booktitle}{\emph{Proceedings of the 2025 ACM/IEEE International
  Conference on Human-Robot Interaction}} \emph{(\bibinfo{series}{HRI '25})}.
  \bibinfo{address}{Melbourne, Australia}, \bibinfo{pages}{869--879}.
\newblock
\urldef\tempurl%
\url{https://doi.org/10.1109/HRI61500.2025.10974218}
\showDOI{\tempurl}


\bibitem[Mur-Labadia et~al\mbox{.}(2026)]%
        {MurLabadia2026VJEPA21}
\bibfield{author}{\bibinfo{person}{Lorenzo Mur-Labadia},
  \bibinfo{person}{Matthew Muckley}, \bibinfo{person}{Amir Bar},
  \bibinfo{person}{Mido Assran}, \bibinfo{person}{Koustuv Sinha},
  \bibinfo{person}{Mike Rabbat}, \bibinfo{person}{Yann LeCun},
  \bibinfo{person}{Nicolas Ballas}, {and} \bibinfo{person}{Adrien Bardes}.}
  \bibinfo{year}{2026}\natexlab{}.
\newblock \bibinfo{title}{{V-JEPA} 2.1: Unlocking Dense Features in Video
  Self-Supervised Learning}.
\newblock
\newblock
\showeprint[arxiv]{2603.14482}~[cs.CV]
\urldef\tempurl%
\url{https://arxiv.org/abs/2603.14482}
\showURL{%
\tempurl}


\bibitem[Nadler(2004)]%
        {Nadler2004RapportNegotiation}
\bibfield{author}{\bibinfo{person}{Janice Nadler}.}
  \bibinfo{year}{2004}\natexlab{}.
\newblock \showarticletitle{Rapport in Negotiation and Conflict Resolution}.
\newblock \bibinfo{journal}{\emph{Marquette Law Review}} \bibinfo{volume}{87},
  \bibinfo{number}{4} (\bibinfo{year}{2004}), \bibinfo{pages}{875--882}.
\newblock
\urldef\tempurl%
\url{https://scholarship.law.marquette.edu/mulr/vol87/iss4/25}
\showURL{%
\tempurl}


\bibitem[Ni et~al\mbox{.}(2022)]%
        {Ni2022SentenceT5}
\bibfield{author}{\bibinfo{person}{Jianmo Ni}, \bibinfo{person}{Gustavo
  Hernandez~Abrego}, \bibinfo{person}{Noah Constant}, \bibinfo{person}{Ji Ma},
  \bibinfo{person}{Keith Hall}, \bibinfo{person}{Daniel Cer}, {and}
  \bibinfo{person}{Yinfei Yang}.} \bibinfo{year}{2022}\natexlab{}.
\newblock \showarticletitle{Sentence-T5: Scalable Sentence Encoders from
  Pre-trained Text-to-Text Models}. In \bibinfo{booktitle}{\emph{Findings of
  the Association for Computational Linguistics: ACL 2022}}.
  \bibinfo{publisher}{Association for Computational Linguistics},
  \bibinfo{address}{Dublin, Ireland}, \bibinfo{pages}{1864--1874}.
\newblock
\urldef\tempurl%
\url{https://doi.org/10.18653/v1/2022.findings-acl.146}
\showDOI{\tempurl}


\bibitem[Nielsen et~al\mbox{.}(2023)]%
        {Nielsen2023UnguidedInteractions}
\bibfield{author}{\bibinfo{person}{Sara Nielsen}, \bibinfo{person}{Mikael~B.
  Skov}, \bibinfo{person}{Karl~Damkj{\ae}r Hansen}, {and}
  \bibinfo{person}{Aleksandra Kaszowska}.} \bibinfo{year}{2023}\natexlab{}.
\newblock \showarticletitle{Using User-Generated YouTube Videos to Understand
  Unguided Interactions with Robots in Public Places}.
\newblock \bibinfo{journal}{\emph{ACM Transactions on Human-Robot Interaction}}
  \bibinfo{volume}{12}, \bibinfo{number}{1}, Article \bibinfo{articleno}{5}
  (\bibinfo{year}{2023}), \bibinfo{numpages}{40}~pages.
\newblock
\urldef\tempurl%
\url{https://doi.org/10.1145/3550280}
\showDOI{\tempurl}


\bibitem[Niemel{\"a} et~al\mbox{.}(2019)]%
        {Niemela2019ShoppingMall}
\bibfield{author}{\bibinfo{person}{Marketta Niemel{\"a}},
  \bibinfo{person}{P{\"a}ivi Heikkil{\"a}}, \bibinfo{person}{Hanna Lammi},
  {and} \bibinfo{person}{Virpi Oksman}.} \bibinfo{year}{2019}\natexlab{}.
\newblock \showarticletitle{A Social Robot in a Shopping Mall: Studies on
  Acceptance and Stakeholder Expectations}.
\newblock In \bibinfo{booktitle}{\emph{Social Robots: Technological, Societal
  and Ethical Aspects of Human-Robot Interaction}}.
  \bibinfo{publisher}{Springer}, \bibinfo{pages}{119--144}.
\newblock
\urldef\tempurl%
\url{https://doi.org/10.1007/978-3-030-17107-0_7}
\showDOI{\tempurl}


\bibitem[Nomura and Kanda(2016)]%
        {NomuraKanda2016RERS}
\bibfield{author}{\bibinfo{person}{Tatsuya Nomura} {and}
  \bibinfo{person}{Takayuki Kanda}.} \bibinfo{year}{2016}\natexlab{}.
\newblock \showarticletitle{Rapport--Expectation with a Robot Scale}.
\newblock \bibinfo{journal}{\emph{International Journal of Social Robotics}}
  \bibinfo{volume}{8}, \bibinfo{number}{1} (\bibinfo{year}{2016}),
  \bibinfo{pages}{21--30}.
\newblock
\urldef\tempurl%
\url{https://doi.org/10.1007/s12369-015-0293-z}
\showDOI{\tempurl}


\bibitem[Pereira et~al\mbox{.}(2024)]%
        {Pereira2024LLMEnjoyment}
\bibfield{author}{\bibinfo{person}{Andr{\'e} Pereira}, \bibinfo{person}{Lubos
  Marcinek}, \bibinfo{person}{Jura Miniota}, \bibinfo{person}{Sofia Thunberg},
  \bibinfo{person}{Erik Lagerstedt}, \bibinfo{person}{Joakim Gustafson},
  \bibinfo{person}{Gabriel Skantze}, {and} \bibinfo{person}{Bahar Irfan}.}
  \bibinfo{year}{2024}\natexlab{}.
\newblock \showarticletitle{Multimodal User Enjoyment Detection in Human-Robot
  Conversation: The Power of Large Language Models}. In
  \bibinfo{booktitle}{\emph{Proceedings of the 26th International Conference on
  Multimodal Interaction}} \emph{(\bibinfo{series}{ICMI '24})}.
  \bibinfo{pages}{469--478}.
\newblock
\urldef\tempurl%
\url{https://doi.org/10.1145/3678957.3685729}
\showDOI{\tempurl}


\bibitem[Radford et~al\mbox{.}(2023)]%
        {Radford2023Whisper}
\bibfield{author}{\bibinfo{person}{Alec Radford}, \bibinfo{person}{Jong~Wook
  Kim}, \bibinfo{person}{Tao Xu}, \bibinfo{person}{Greg Brockman},
  \bibinfo{person}{Christine McLeavey}, {and} \bibinfo{person}{Ilya
  Sutskever}.} \bibinfo{year}{2023}\natexlab{}.
\newblock \showarticletitle{Robust Speech Recognition via Large-Scale Weak
  Supervision}. In \bibinfo{booktitle}{\emph{Proceedings of the 40th
  International Conference on Machine Learning}}
  \emph{(\bibinfo{series}{Proceedings of Machine Learning Research},
  Vol.~\bibinfo{volume}{202})}. \bibinfo{publisher}{PMLR},
  \bibinfo{pages}{28492--28518}.
\newblock
\urldef\tempurl%
\url{https://proceedings.mlr.press/v202/radford23a.html}
\showURL{%
\tempurl}


\bibitem[Santana et~al\mbox{.}(2025)]%
        {Santana2025SpeechToJoy}
\bibfield{author}{\bibinfo{person}{Ricardo Santana}, \bibinfo{person}{Bahar
  Irfan}, \bibinfo{person}{Erik Lagerstedt}, \bibinfo{person}{Gabriel Skantze},
  {and} \bibinfo{person}{Andr{\'e} Pereira}.} \bibinfo{year}{2025}\natexlab{}.
\newblock \showarticletitle{Speech-to-Joy: Self-Supervised Features for
  Enjoyment Prediction in Human-Robot Conversation}. In
  \bibinfo{booktitle}{\emph{Proceedings of the 27th International Conference on
  Multimodal Interaction}} \emph{(\bibinfo{series}{ICMI '25})}.
  \bibinfo{pages}{238--248}.
\newblock
\urldef\tempurl%
\url{https://doi.org/10.1145/3716553.3750747}
\showDOI{\tempurl}


\bibitem[Sinha and Cassell(2015)]%
        {SinhaCassell2015WeClick}
\bibfield{author}{\bibinfo{person}{Tanmay Sinha} {and} \bibinfo{person}{Justine
  Cassell}.} \bibinfo{year}{2015}\natexlab{}.
\newblock \showarticletitle{We Click, We Align, We Learn: Impact of Influence
  and Convergence Processes on Student Learning and Rapport Building}. In
  \bibinfo{booktitle}{\emph{Proceedings of the 1st Workshop on Modeling
  INTERPERsonal SynchrONy And infLuence}} \emph{(\bibinfo{series}{INTERPERSONAL
  '15})}. \bibinfo{publisher}{ACM}, \bibinfo{pages}{13--20}.
\newblock
\urldef\tempurl%
\url{https://doi.org/10.1145/2823513.2823516}
\showDOI{\tempurl}


\bibitem[Tickle-Degnen and Rosenthal(1990)]%
        {TickleDegnenRosenthal1990NatureRapport}
\bibfield{author}{\bibinfo{person}{Linda Tickle-Degnen} {and}
  \bibinfo{person}{Robert Rosenthal}.} \bibinfo{year}{1990}\natexlab{}.
\newblock \showarticletitle{The Nature of Rapport and Its Nonverbal
  Correlates}.
\newblock \bibinfo{journal}{\emph{Psychological Inquiry}} \bibinfo{volume}{1},
  \bibinfo{number}{4} (\bibinfo{year}{1990}), \bibinfo{pages}{285--293}.
\newblock
\urldef\tempurl%
\url{https://doi.org/10.1207/s15327965pli0104_1}
\showDOI{\tempurl}


\bibitem[Wei et~al\mbox{.}(2025)]%
        {Wei2025AdversarialRapport}
\bibfield{author}{\bibinfo{person}{Wenqing Wei}, \bibinfo{person}{Sixia Li},
  \bibinfo{person}{Candy~Olivia Mawalim}, \bibinfo{person}{Xiguang Li},
  \bibinfo{person}{Kazunori Komatani}, {and} \bibinfo{person}{Shogo Okada}.}
  \bibinfo{year}{2025}\natexlab{}.
\newblock \showarticletitle{Influence of Personality Traits and Demographics on
  Rapport Recognition Using Adversarial Learning}.
\newblock \bibinfo{journal}{\emph{Multimodal Technologies and Interaction}}
  \bibinfo{volume}{9}, \bibinfo{number}{3}, Article \bibinfo{articleno}{18}
  (\bibinfo{year}{2025}).
\newblock
\urldef\tempurl%
\url{https://doi.org/10.3390/mti9030018}
\showDOI{\tempurl}


\bibitem[Wei et~al\mbox{.}(2022)]%
        {Wei2022DialogueExchangeImpression}
\bibfield{author}{\bibinfo{person}{Wenqing Wei}, \bibinfo{person}{Sixia Li},
  {and} \bibinfo{person}{Shogo Okada}.} \bibinfo{year}{2022}\natexlab{}.
\newblock \showarticletitle{Investigating the Relationship Between Dialogue and
  Exchange-Level Impression}. In \bibinfo{booktitle}{\emph{Proceedings of the
  2022 International Conference on Multimodal Interaction}}
  \emph{(\bibinfo{series}{ICMI '22})}. \bibinfo{pages}{359--367}.
\newblock
\urldef\tempurl%
\url{https://doi.org/10.1145/3536221.3556602}
\showDOI{\tempurl}


\bibitem[Wei et~al\mbox{.}(2021)]%
        {Wei2021Satisfaction}
\bibfield{author}{\bibinfo{person}{Wenqing Wei}, \bibinfo{person}{Sixia Li},
  \bibinfo{person}{Shogo Okada}, {and} \bibinfo{person}{Kazunori Komatani}.}
  \bibinfo{year}{2021}\natexlab{}.
\newblock \showarticletitle{Multimodal User Satisfaction Recognition for
  Non-task Oriented Dialogue Systems}. In \bibinfo{booktitle}{\emph{Proceedings
  of the 2021 International Conference on Multimodal Interaction}}
  \emph{(\bibinfo{series}{ICMI '21})}. \bibinfo{pages}{586--594}.
\newblock
\urldef\tempurl%
\url{https://doi.org/10.1145/3462244.3479928}
\showDOI{\tempurl}


\bibitem[Yano et~al\mbox{.}(2026)]%
        {Yano2026Breaking}
\bibfield{author}{\bibinfo{person}{Yuga Yano}, \bibinfo{person}{Yuki Okafuji},
  \bibinfo{person}{Ryo Miyoshi}, \bibinfo{person}{Sanae Yamashita}, {and}
  \bibinfo{person}{Yoshiki Ohira}.} \bibinfo{year}{2026}\natexlab{}.
\newblock \bibinfo{title}{Breaking the 15\% Barrier: A Real-World Data-Driven
  System for Proactive Social Robot Triggered by User Nonverbal Cues}.
\newblock
\newblock
\showeprint[arxiv]{2607.11633}~[cs.RO]
\urldef\tempurl%
\url{https://arxiv.org/abs/2607.11633}
\showURL{%
\tempurl}
\newblock
\shownote{Accepted at the 2026 IEEE/RSJ International Conference on Intelligent
  Robots and Systems (IROS 2026)}.


\end{thebibliography}

\end{document}